\documentclass[11pt]{article}

\usepackage{amsmath}
\usepackage{amssymb,mathrsfs}

\usepackage{graphicx}

\usepackage{cite}

\usepackage{color}

\usepackage{xr}
\def\X{{\mathscr{X}}}

\usepackage[labelfont=bf,labelsep=period,justification=raggedright]{caption}

\makeatletter
\renewcommand{\@biblabel}[1]{\quad#1.}
\makeatother

\date{}

\begin{document}


\begin{flushleft}
{\Large
\textbf{An HMM-based Comparative Genomic Framework for Detecting Introgression in Eukaryotes}
}
\\
Kevin J. Liu$^{1,2,\ast}$, 
Jingxuan Dai$^{1}$,
Kathy Truong$^{1}$, 
Ying Song$^{2}$,
Michael H. Kohn$^{2}$,
Luay Nakhleh$^{1, 2,\ast}$
\\
\bf{1} Department of Computer Science, Rice University, Houston, TX, United States of America
\\
\bf{2} Department of Ecology and Evolutionary Biology, Rice University, Houston, TX, United States of America
\\
$\ast$ E-mail: kl23@rice.edu, nakhleh@rice.edu
\end{flushleft}

\section*{Abstract}
One outcome of interspecific hybridization and subsequent effects of evolutionary 
 forces is introgression, which is  the integration of genetic material from one species 
 into the genome of an individual in another species. The evolution of several groups of 
 eukaryotic species has involved hybridization, and cases of adaptation through 
 introgression have been already established. In this work, we report on a new 
 comparative genomic framework for detecting introgression in genomes, called PhyloNet-HMM, which combines 
 phylogenetic networks, that capture reticulate evolutionary relationships among genomes,
 with hidden Markov models (HMMs), that capture dependencies within genomes. 
 A novel aspect of our work is that it also accounts for incomplete lineage sorting and 
  dependence across loci.
 Application of 
 our model to variation data from chromosome 7 in the mouse ({\em Mus musculus domesticus}) genome detects a recently
reported adaptive introgression event involving the rodent poison resistance gene {\em Vkorc1}, in addition to 
other newly detected introgression regions. Based on our analysis, it is estimated that about 12\% of all sites within
chromosome 7 are of introgressive origin (these cover about 18 Mbp of chromosome 7, and over 300 genes). Further, our model 
detects no introgression in two negative control data sets. Our work provides a powerful framework for systematic 
analysis of introgression while simultaneously accounting for dependence across sites, point mutations, recombination, 
 and ancestral polymorphism. 

\section*{Author Summary}
Hybridization is the mating between individuals from two different species. While hybridization introduces 
genetic material into a host genome, this genetic material may be transient and is purged from the population within a few generations 
after hybridization. However, in other cases, the introduced 
genetic material persists in the population---a process known as introgression---and  can have significant evolutionary implications.
In this paper, we introduce a novel method for detecting introgression in genomes using a comparative genomic approach. 
The method scans multiple aligned genomes for signatures of introgression by incorporating phylogenetic networks and hidden Markov models. The method allows for teasing apart true signatures of introgression from spurious ones that arise due to 
population effects and resemble those of introgression. Using the new method, we analyzed three  sets of variation data 
from chromosome 7 in mouse genomes. The method detected previously reported introgressed regions as well as new ones in 
one of the data sets. In the other two data sets, which were selected as negative controls, the method detected no introgression. 
Our method enables systematic comparative analyses of genomes where introgression is suspected, and can work with genome-wide data. 

 \section*{Introduction}

Hybridization is the mating between species that can result in the
transient or permanent transfer of genetic variants from one species
to another. The latter outcome is referred to as introgression.  Mallet \cite{Mallet05}  recently 
 estimated that ``at least 
25\% of plant species and 10\% of animal species, mostly the youngest species, 
are involved in hybridization and potential introgression with other species." 
Introgression can be neutral and 
 go 
 unnoticed in terms of
phenotypes but can also be adaptive and affect phenotypes.
  Recent
examples of adaptation through hybridization include resistance to
rodenticides in mice \cite{Song20111296} and mimicry in butterflies
\cite{Heliconius2012}. Detecting regions with signatures of
 introgression in eukaryotic genomes is of great interest, given the 
 consequences of introgression in evolutionary  biology, speciation, 
 biodiversity, and conservation \cite{Mallet05}. With the increasing availability of 
 genomic data, it is imperative to develop techniques that detect genomic 
 regions of introgressive descent. 

\begin{figure}[!ht]
\begin{center}
\includegraphics[width=6.5in]{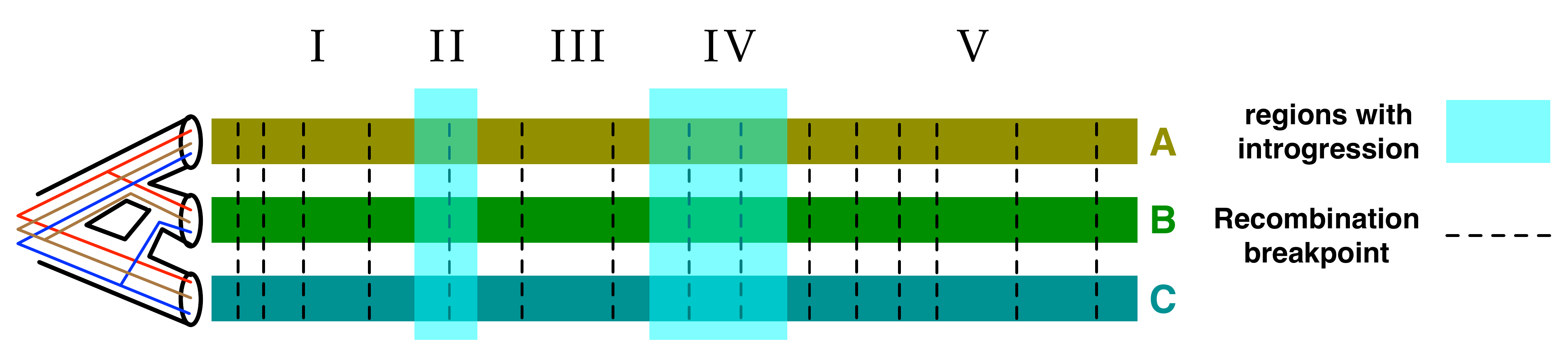} 
\caption{\small{\bf Evolutionary and genomic views of three genomes involving introgression.} 
  Hybridization between species B and C results in individuals of species B with genomes that 
 are mosaics with regions of ``vertical" descent from B and others of introgressive descent from C. 
 Walking along the genomes from left to right, local gene genealogies are observed, and when a 
 recombination breakpoint is crossed, the local genealogy changes. Switching of local gene 
 genealogies of unlinked (broken by recombination) loci is known as incomplete lineage sorting (ILS). 
 Further, the walk enters regions of introgressive descent (II and IV), where the genealogies switch 
 due to hybridization. The complexity of the model stems from the co-occurrence of ILS and introgression,
 and the need to tease them apart. Within the phylogenetic network of the species (leftmost), 
 three possible gene genealogies are shown: one that agrees with how species split and diverged (red), one that 
 is reflective of the introgression event (blue), and another that is a signature of ILS (brown).  \label{fig3}}
\end{center}
\end{figure}

Let us consider an evolutionary scenario where two speciation events result in 
three extant species A, B, and C, with A and B sharing a most recent common ancestor. Further, some time after 
the splitting of A and B, hybridization occurs between B and C (that is, sexual reproduction of 
individuals from these two species). This scenario is depicted by the phylogenetic network 
in Fig.~\ref{fig3}. 
 Immediately upon hybridization, approximately half of
the hybrid individual's genome comes from an individual in species
$B$, whereas the remainder comes from an individual in species
$C$. However, in homoploid hybridization, where the hybrid offspring
has the same ploidy level as the two parental species, hybridization
is often followed by back-crossing (further mating between the hybrid
population and either of the two parental populations). Repeated
back-crossing, followed by the effects of genetic drift and natural
selection, results in genomes in the hybrid individuals that are
mosaics of genomic material from the two parental species, yet not
necessarily with a 50-50 composition. Thus, detecting introgressed 
regions requires scanning across the genome and looking for signals of 
introgression. 

In a comparative framework, detecting introgressed regions can be
achieved by evolutionary analysis of genomes from the parental
species, as well as genomes from introgressed individuals. In such an
analysis, a walk across the genomes is taken, and local phylogenies are 
(or, genealogies) are inspected; incongruence between two local phylogenies 
 can be taken as a signal of introgresison \cite{Maddison97}.  However, in reality, the analysis is
more involved than this, owing to potentially confounding signal
produced by several factors, major of which is incomplete lineage sorting (ILS). 
 As recombination breaks linkage across loci in the genome, the result is independent loci 
that may have different genealogies by chance, which is known as ILS.  
 ILS is 
 common
to several groups of eukaryotic taxa where species diverged
recently and not enough time has elapsed for all genomic loci to
completely sort, resulting in a scenario 
where introgression and
ILS effects need to be distinguished \cite{Green07052010,ErikssonManica12,Staubach12,Heliconius2012,MoodyRieseberg12}.
 Fig.~\ref{fig3} illustrates this issue, where local genealogies across recombination breakpoints differ due to ILS, but 
 also differ inside vs. outside introgressed regions.  
 While other factors, such as gene duplication and loss \cite{Nakhleh2013}, could potentially add to the complexity 
 of the phylogenetic and genomic patterns, we focus here on introgression and ILS. 
%
%

Recently,
new methods were proposed to detect introgression in the presence of ILS.
Durand {\em et al.}'s $D$ statistic allows for a sliding-window analysis of three-taxon 
 data sets, while accounting for introgression and ancestral polymorphism \cite{Durand01082011}.
 However, this statistic assumes an infinite-sites model and independence across loci. 
 Yu {\em et al.} \cite{Yu2012} proposed a new statistical model for the likelihood of a species 
 phylogeny model, given a set of gene genealogies, accounting for both ILS and introgression. 
 However, this model does not work directly from the sequences; rather, it assumes that gene genealogies 
 have been estimated, and computations are based on these estimates. Further, the model assumes 
 independence across loci. Of great relevance to our work here is a  group of hidden Markov model
(HMM) based techniques were introduced recently for analyzing genomic data in the presence of 
 recombination and ILS \cite{Hobolth2007,Dutheil2009,Mailund2011}; however, these methods do not 
 account for introgression. 
 A recent extension \cite{Mailund2012} was devised to investigate the effects of population structure.
 
In this paper, we devise a novel model based on integrating
phylogenetic networks with hidden Markov models (HMMs). The phylogenetic
network component of our model captures the relatedness across genomes (including
point mutation, recombination, ILS, and introgression), and the HMM
component captures dependence across sites and loci within each
genome.
Using dynamic programming algorithms \cite{Rabiner89atutorial} paired with a multivariate
optimization heuristic \cite{Brent1973}, the model can be trained on genomic data, and allows 
for the identification of genomic regions of introgressive descent. We applied our model to chromosome 
7 genomic variation data from three mouse data sets. Our analysis recovered an introgression event 
involving the rodenticide resistance gene {\em VKORC1}, which was recently reported in the 
literature \cite{Song20111296}. Further, our analysis indicates that about 12\% of sites within chromosome 7 
are in fact of introgressive origin, which is a novel finding. Further, when applied to two negative control data sets,
 our model did not detect any introgression, further attesting to its robustness. 
 Our model will enable new analyses of eukaryotic data sets where introgression is suspected, and will 
further help shed light on the Tree of Life---or, Network of Life. 
%
%

\section*{Materials and Methods}

\subsection*{Sample selection and sequence data}


Our study utilizes eight mice that were either newly sampled or from previous
publications. Details
for the eight mice are listed in Table \ref{samples} in the Appendix.
Newly sampled mice were obtained as part of a tissue sharing agreement 
between Rice University and Stefan Endepols at Environmental Science, Bayer CropScience AG, D-40789 Monheim, Germany 
and Dania Richter and Franz-Rainer Matuschka at Division of Pathology, Department of Parasitology, 
Charit\'{e}-Universit\"{a}tsmedizin, D-10117 Berlin, Germany
(reviewed and exempted by Rice University IACUC).

 The {\em M. m. domesticus} data set was constructed as follows. We included a wild {\it M. m. domesticus} sample from Spain, part of the sympatry region between 
{\it M. m. domesticus} and {\it M. spretus}.
To help maximize genetic differences as part of the design
goals of our pipeline, we also selected a
``baseline'' {\it M. m. domesticus} sample that originated
from a region as far from the sympatry region as possible.
Thus, we obtained an {\it M. m. domesticus} sample
from Georgia.
 We utilized two {\it M. spretus} samples.
The samples came from different parts of the sympatry
region in Spain.

 The reference strain control data set contained two mice from the C57BL/6J strain and the above two {\em M. spretus} samples. 
 The {\em M. m. musculus} control data set contained two wild
{\it M. m. musculus} samples from China and the above two {\em M. spretus} samples. 

The Mouse Diversity Array was used to obtain all sequence data in our study \cite{Yang2009}.
 Sequence data for all other mice were 
obtained from past studies \cite{Yang2011,Didion2012,Guenet200324}. 
 Since the probe sets in these studies
differed slightly, we used the intersection of the probe sets
in our study.
 A total of 535,988 probes were used.

We genotyped all raw sequencing reads using MouseDivGeno version 1.0.4 \cite{Didion2012}.
We utilized a threshold for genotyping confidence scores
of 0.05.
We phased all genotypes into haplotypes and imputed bases for missing data using fastPHASE \cite{Scheet2006629}.
Less than 15.1\% of genotype calls were heterozygous or missing and thus affected by the fastPHASE analysis.
The genotyping and phasing analyses were performed with a larger superset of samples. 
The additional samples consisted of
the 362 samples used in \cite{Didion2012} that were otherwise not used in our study.
After genotyping and phasing was completed, we thereafter used only the samples listed in Table \ref{samples} in the Appendix.


Genomic coordinates and annotation in our study were based on the 
MGSCv37 reference mouse genome (GenBank accession GCA\_000001635.1).
MouseDivGeno also makes use of data from the MGSCv37 reference genome.

%
%


\subsection*{The PhyloNet-HMM model: A simple case first}
\begin{figure}[!ht]
\begin{center}
\begin{tabular}{ccc}
\includegraphics[width=1.2in]{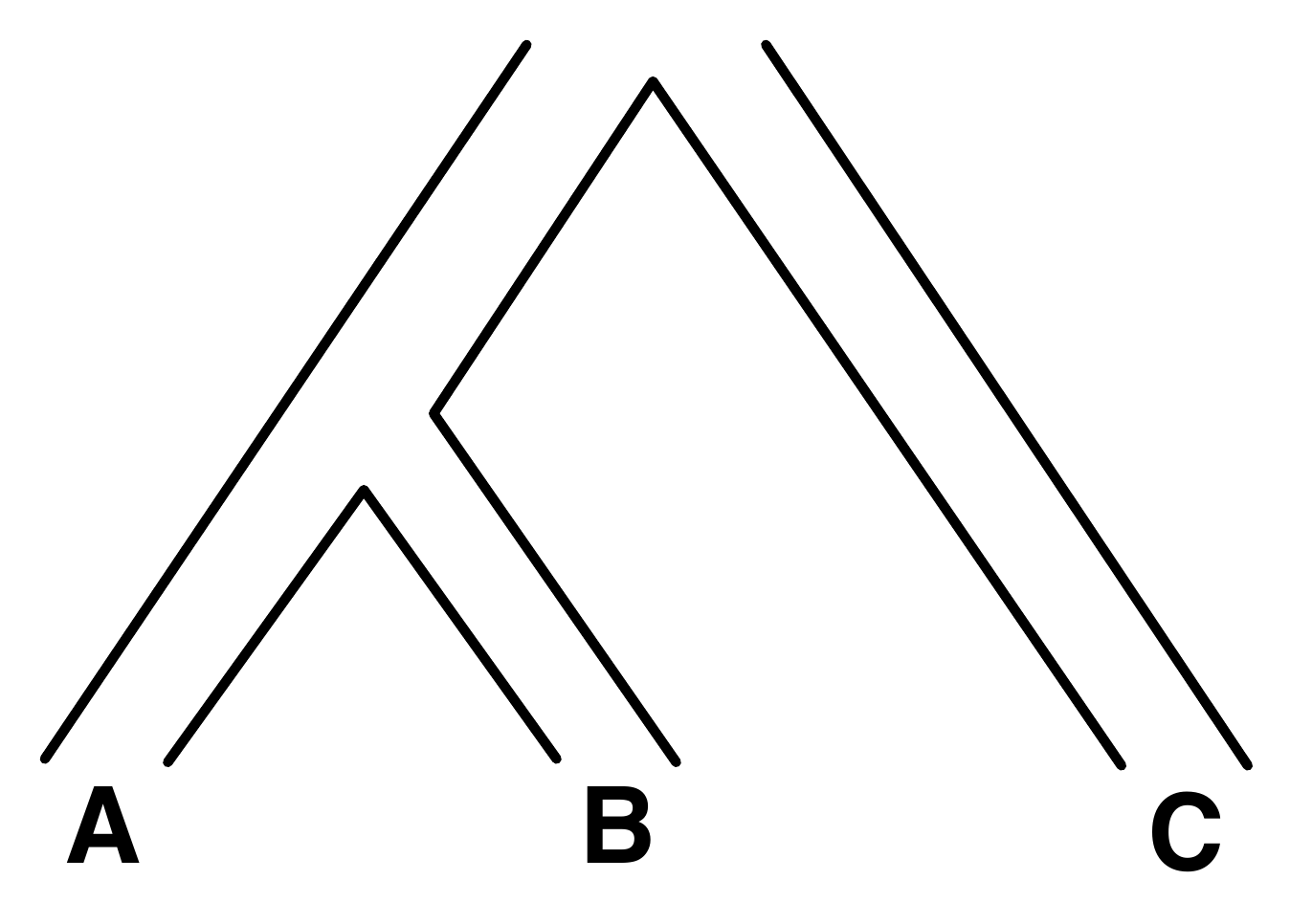} &
\includegraphics[width=1.2in]{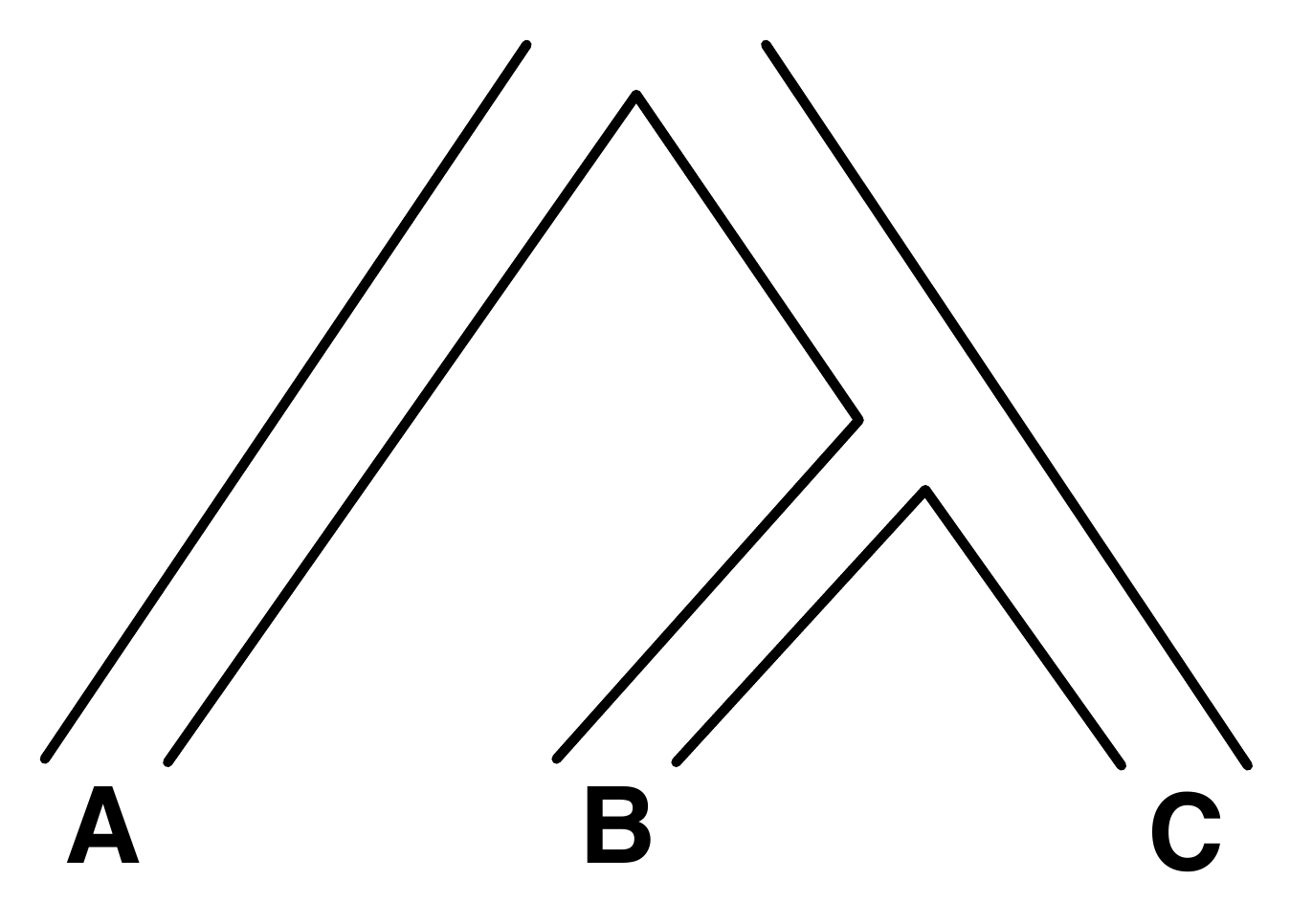} &
\includegraphics[width=3in]{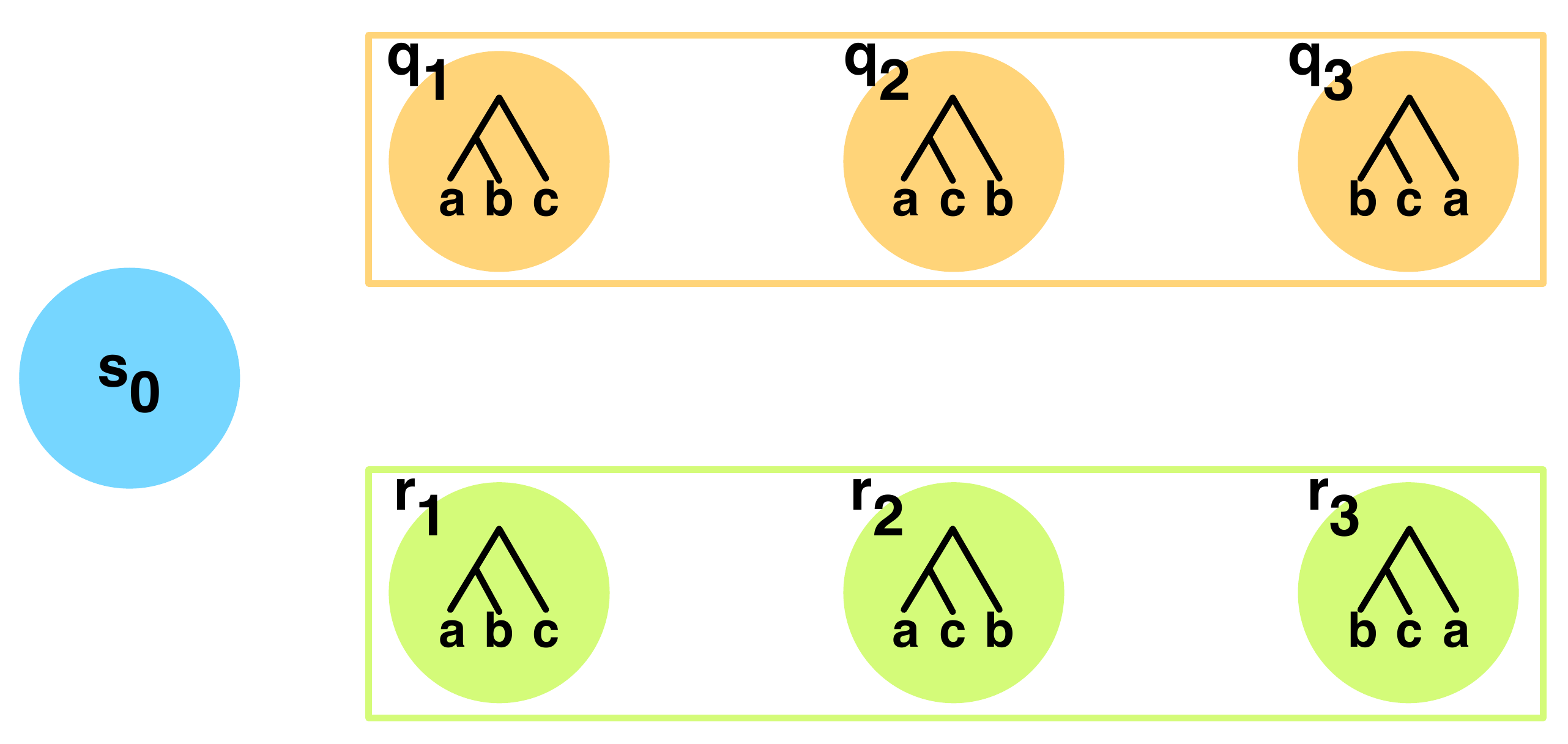} \\
(a)  & (b) & (c)  \\
\end{tabular}
\caption{\small{\bf The two parental
    trees that represent the phylogenetic network of Fig.~\ref{fig3} and the corresponding PhyloNet-HMM.} genomic regions that are not of 
    introgressive descent evolve within the branches of the parental tree in (a), whereas genomic 
    regions of introgressive descent evolve within the branches of the parental tree in (b). Regardless of which of the two 
    parental trees a genomic region evolves, the region might still have phylogenetic switching, which is captured by 
    different gene genealogies within the branches of each parental tree. 
    (c) The corresponding PhyloNet-HMM (only states are shown). The three $q$ states correspond to genomic regions
  whose evolution follows the parental tree in (a), and there is a state for each of the three possible gene genealogies. The  three $r$ states
  correspond to genomic regions whose evolution follows the parental
  tree in (b), and there is a state  for each of the three possible gene genealogies. $s_0$ is the start state. See text for emission and transition probabilities. \label{fig2}}
\end{center}
\end{figure}

 Let us revisit the scenario of Fig.~\ref{fig3}. In this specific case, where only one individual is samples from each of the three 
 species, each local gene genealogy (or, gene tree), 
 has evolved within one of two ``parental trees" that represent the phylogenetic network \cite{MengKubatko08}; see 
 Fig.~\ref{fig2}(a-b) for an illustration. 
To account for this case, we propose a  hidden Markov model
(HMM) for modeling the evolution of the three genomes.  The HMM for
this simple case would consist of 7 states: a start state $s_0$, and
six additional states: $q_i$ ($1 \leq i \leq 3$), corresponding to gene possible gene genealogies within 
one parental tree, and $r_i$ ($1 \leq i \leq 3$), corresponding to gene possible gene genealogies within 
the other parental tree (see Fig.~\ref{fig2}(c)). We denote by $g(q_i)$ and $g(r_i)$ the gene trees to which 
states $q_i$ and $r_i$ correspond, respectively (these gene trees are shown within the states in 
Fig.~\ref{fig2}(c)). 

In this model, transition between two $q$ states or two $r$ states corresponds to switching across recombination 
 breakpoints. The probabilities of such transitions have to do with population parameters (e.g., population size, 
 recombination rates, etc.).  Transition from a $q$ state to an $r$ state indicates entering a introgressed region, while transition from 
 an $r$ state to a $q$ state indicates exiting an introgressed region. The probabilities of such transitions have to do, in addition, with 
 introgression and evolutionary forces (back-crossing, selection, etc.). Each state emits a triplet of letters that corresponds 
 to a column in the three-genome sequence alignment. The probability of emitting such a triplet can be computed using a 
 standard phylogenetic substitution model \cite{Fel04}. 
 
Following the approaches of \cite{Hobolth2007,Westesson2009},
the transition probabilities in our model do
not represent parameters in an explicit evolutionary model of
recombination and introgression.
Our choice was made to ease
analytical representation and to permit tractable computational inference.
We contrast our choice with alternative approaches:
examples include (in
order of increasing tractability of computational inference
at the cost of more simplifying assumptions) methods incorporating
the coalescent-with-recombination model \cite{Husmeier2001},
the sequentially Markovian coalescent-with-recombination model \cite{Dutheil2009}
(which adds the single assumption that coalescence cannot
occur between two lineages that do not share ancestral genetic
material),
and the discretized sequentially Markovian coalescent-with-recombination model
\cite{Li2011} (which additionally
discretizes time).

Assume that the probability of a site (or, locus) in the genome of B being introgressed is $\gamma$, then 
we follow the model of \cite{Yu2012}, and use this parameter to constrain the transition probabilities. 
Assume a site is emitted by state $q_1$ and consider the next site. If it is not in an introgressed region, then 
the HMM should stay in the $q$ states, with probability $1-\gamma$, and if it is in an introgressed region,  
it should switch, with probability $\gamma$, to an $r$ state. Thus, the transition probability from $q_1$ to 
any other $q_i$ ($1 \leq i \leq 3$) state is $(1-\gamma) z(q_i)$ and to any $r_i$ ($1 \leq i \leq 3$) state 
is $\gamma z(r_i)$, where $z(q_i)$ is the probability of the gene tree $g(q_i)$ given the parental tree in 
Fig.~\ref{fig2}(a), and $z(r_i)$ is the probability of the gene tree $g(r_i)$ given the parental tree in 
Fig.~\ref{fig2}(b). The $z$ quantities are computable under the coalescent using the technique of \cite{DegnanSalter05}. 

%

If we denote by $S$ the set $\{q_1,q_2,q_3,r_1,r_2,r_3\}$ of (non-start) states, then 
 a transition from the start state $s_0$ to a state $s \in S$ occurs according to the
the normalized gene tree probability 
$$t_{s} = \frac{z(s)}{\sum\limits_{s' \in S} z(s')}.$$
 For $s \in S$, let $n_s = (1-\gamma)z(s)$ and $i_s = \gamma z(s)$. 
 Then, the full transition probability matrix, with rows labeled $s_0, q_1, q_2, q_{3}, r_1, r_2, r_{3}$ and
similarly for columns, is

\[ \left[ \begin{array}{ccccccc}
0 & t_{q_1} & t_{q_2} &t_{q_3} & t_{r_1} & t_{r_2} & t_{r_3}   \\
0 &  n_{q_1} & n_{q_2} &n_{q_3} &  i_{r_1} & i_{r_2} & i_{r_3} \\
0 &  n_{q_1} & n_{q_2} &n_{q_3} &  i_{r_1} & i_{r_2} & i_{r_3}  \\
0 &  n_{q_1} & n_{q_2} &n_{q_3} &  i_{r_1} & i_{r_2} & i_{r_3}  \\
0 & i_{q_1} & i_{q_2} & i_{q_3} & n_{r_1} & n_{r_2} & n_{r_3} \\
0 & i_{q_1} & i_{q_2} & i_{q_3} & n_{r_1} & n_{r_2} & n_{r_3} \\
0 & i_{q_1} & i_{q_2} & i_{q_3} & n_{r_1} & n_{r_2} & n_{r_3} \\
\end{array}
\right]
\]


Given that 
$$\sum_{i=1}^3 z(q_i) = \sum_{i=1}^3 z(r_i) = 1,$$
it follows that the entries in each row of the matrix add up to $1$. Further, the HMM always starts in state $s_0$; that is 
the initial state probability distribution is given by $1$ for state $s_0$ and $0$ for every other state. 

%
%
%
%
%
%
%

Once in a  state $s \in S$,
the HMM emits an observation $O_t$ where $1 \leq t \leq K$.
Emissions occur according to a substitution model $\theta$ 
(we used
the generalized time-reversible (GTR) model \cite{GTR}),
yielding the emission probability

\begin{eqnarray}
e_{s, \theta}(O_t) & = & P[O_t | x_t = s, \theta] \nonumber \\
 & = & P[O_t | g(s), b_{g(s)}, \theta ] \nonumber,
\end{eqnarray}
where $b_{g(s)}$ are the branch lengths of the gene tree associated with state $s_i$. 






\subsection*{The PhyloNet-HMM model: The general case}
Modeling a phylogenetic network in terms of a set of parental trees
fails for most cases \cite{YuEtAl11}. For example, if two individuals
are sampled from species B in Fig.~\ref{fig3}, then one allele of a
certain locus in one individual may trace the left parent (to C),
while another allele of the same locus but in the other individual may
trace the right parent (to A). Neither of the two parental trees in 
Fig.~\ref{fig2} can
capture this case. Similarly, if one individual is sampled per
species, but multiple introgression events occur or divergence events
follow the introgression, the concept of parental trees collapses
\cite{Yu2012}. 

To deal with the
general case---where multiple introgressions could occur, multiple
individuals could be sampled, and introgressed species might split and
diverge (and even hybridize again later)---we propose the following approach 
that is based on MUL-trees \cite{Yu2012}.

\begin{figure}[ht] 
\centerline{\includegraphics[width=0.5\textwidth]{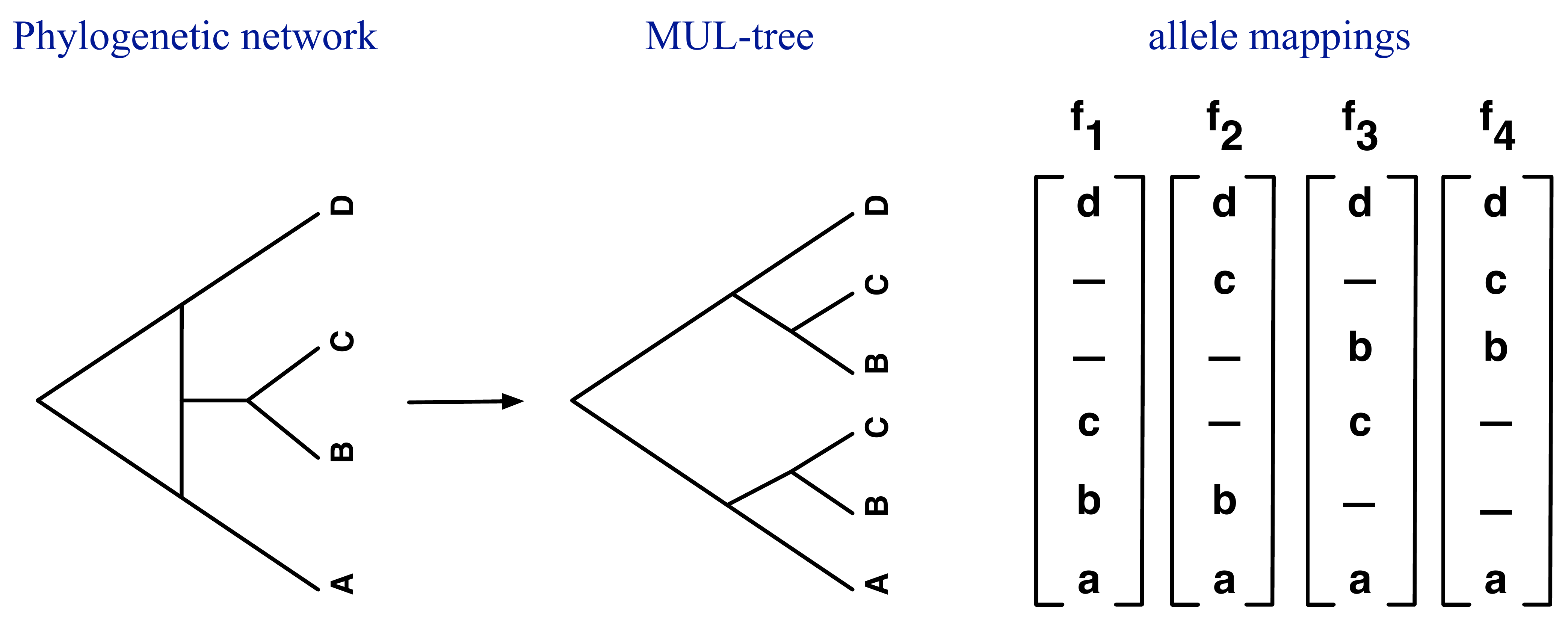}}
\caption{\small {\bf From a phylogenetic network to a MUL-tree.} Illustration of the conversion from a phylogenetic network to a MUL-tree, along with all  allele mappings associated with the case in which single alleles $a$, $b$, $c$ and $d$ were sampled from each of the four species $A$, $B$, $C$ and $D$, respectively. \label{fig:Net2Tree}}
\end{figure}

The basic idea of the method is to convert the phylogenetic network $N$ into a \textit{MUL-tree} $T$ and then make use of some existing techniques to complete the computation on $T$ instead of on $N$. A MUL-tree \cite{HuberOxelman06} is a tree whose leaves are not uniquely labeled by a set of taxa. Therefore, alleles of individuals sampled from one 
 species, say $x$, can map to any of the leaves in the MUL-tree $T$ that are labeled by $x$. For network $N$ on taxa $\X$, we denote by $A_x$ the set of alleles sampled from species $x$ ($x \in \X$), and by $c_x$ the set of leaves in $T$ that are labeled by species $x$. Then an \textit{allele mapping} is a function $f:(\cup_{x \in \X}A_x) \rightarrow (\cup_{x \in \X}c_x)$ such that if $f(a) = d$, and $d \in c_x$, then $a \in A_x$ \cite{YuDegnanNakhleh12}. Fig. \ref{fig:Net2Tree} shows an example of converting a phylogenetic network into a MUL-tree along with all allele mappings when a single allele is sampled per species.
The branch lengths and inheritance probabilities $\gamma$ are transferred from the phylogenetic network to the MUL-tree 
in a straightforward manner (see \cite{Yu2012} for details). 

Now, two changes to the PhyloNet-HMM given for the simple case above are required.
While in the simple case above, we used two classes of states (the $q$
and $r$ states), in the general case, the PhyloNet-HMM will contain $k$ classes of states, where
$k$ is the number of all possible  allele mappings. As above, the
transitions within a class of states corresponds to local phylogeny
switching due to recombination and ILS, whereas transitioning between
classes corresponds to introgression breakpoints. Second, the transition probabilities 
 are now computed
using the method of \cite{Yu2012}, since the methods of
\cite{DegnanSalter05,Wu12} are not applicable to MUL-trees. 



\subsection*{Learning the model and conducting inference}

We adopted an expectation-maximization (EM) approach to infer
model parameters $\lambda$ that
maximize the  likelihood of the model 
 $P[O_1, \ldots, O_K | \lambda]$. Here, $\lambda$ consists of the (1) parental tree branch lengths, 
  (2) the gene genealogy branch lengths,
  (3) the substitution model parameters $\theta$, 
  and (4) the
parental tree switching probability, $\gamma$. Notice that that $z()$ values are 
 completely determined by the parental tree branch lengths and gene tree topology; 
 hence, they are not free parameters in this model. 
 
 The standard forward and backward
algorithms \cite{Rabiner89atutorial} were used to compute
the model likelihood for fixed $\lambda$.
We used Brent's method \cite{Brent1973} as a 
univariate optimization heuristic during each E-M iteration.
To reduce overfitting during optimization, branch length parameters
were optimized for each topologically distinct parental tree,
and similarly for each topologically distinct unrooted gene genealogy
(since we use a reversible substitution model).  States therefore
``shared'' branch length parameters based on topological equivalence
of parental trees and gene genealogies.

To evaluate the effectiveness of our optimization heuristic, 
we utilized different starting points for our E-M search.
We found that our heuristics were robust to 
the choice of starting point since the searches all converged to the same solution (data not shown).
We found that the choice of starting point only affected search time.

After model parameter values were inferred using
the E-M heuristic, Viterbi's algorithm \cite{Rabiner89atutorial} 
was used to compute optimal paths and, thus, annotations of the genomes. 
More formally, Viterbi algorithm computes the path of states $\pi$ such that 
$$\pi \leftarrow {\rm argmax}_{\pi'} {\bf P}(x,\pi'),$$
where $x$ is the sequence alignment.

Further, the forward and backward algorithms can be used to conduct posterior decoding and 
 assess confidence for the states on a path $\pi$: 
 $${\bf P}(\pi_i = k|x) = \frac{f_k(i)b_k(i)}{{\bf P}(x)},$$
 where $f_k(i)$ is the probability of the observed sequence alignment up to and include column $i$,
  requiring that $\pi_i=k$ (computable with the forward algorithm); $b_k(i)$ is the probability of the last $L-i$ 
  columns ($L$ is the total number of columns in the alignment), requiring that $\pi=k$ (computable with the 
  backward algorithm); and, ${\bf P}(x)$ is the probability of the alignment (computable with either the forward or 
  backward algorithms). 
  
  We will show in the Results section the application of both Viterbi's algorithm and the posterior decoding 
  in detecting introgression in genomes.

\section*{Results/Discussion}
\begin{figure}[!ht]
\begin{center}
\includegraphics[width=4in]{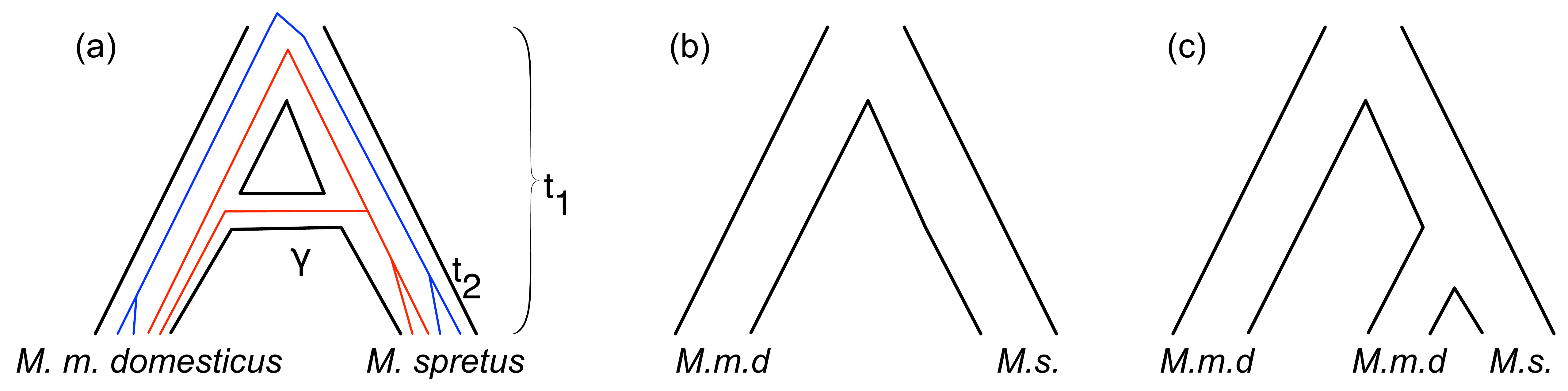}
\caption{\small{\bf The phylogenetic network
used in our analyses and the two parental trees.} The phylogenetic network (a) captures 
introgression from {\em M. spretus} to {\em M. m. domesticus}. The red and blue lines illustrate two 
possible gene genealogies involving no introgression (blue) and introgression (red). The parental tree in (b) captures  
genomic regions with no introgression, while the parental tree in (c) captures genomic regions 
of introgressive descent. 
\label{phmm-parental-trees}
}
\end{center}
\vspace{-.3in}
\end{figure}

 We applied the PhyloNet-HMM framework to detect introgression in chromosome 7 in three sets 
 of mice, as described above. Each data set consisted of two individuals from {\em M. m. domesticus} 
 and two individuals from {\em M. spretus}. Thus the phylogenetic network is very simple, and has only 
 two leaves, with a reticulation edge from {\em M. spretus} to {\em M. m. domesticus}; see Fig.~\ref{phmm-parental-trees}(a). 
Similarly to the example in Fig. \ref{fig3} and \ref{fig2}, 
the evolution of lineages within the species network
can be equivalently captured by the set of parental trees
in Fig. \ref{phmm-parental-trees}(b-c).
 Since in each data set we have four genomes, there are 15 possible rooted gene trees on four taxa. Therefore, for each 
data set, our model consisted of 15 $q$ states, 15 $r$ states, and one start state $s_0$, for a total of 31 states.

\begin{figure}[!ht]
\begin{center}
\includegraphics[width=5in]{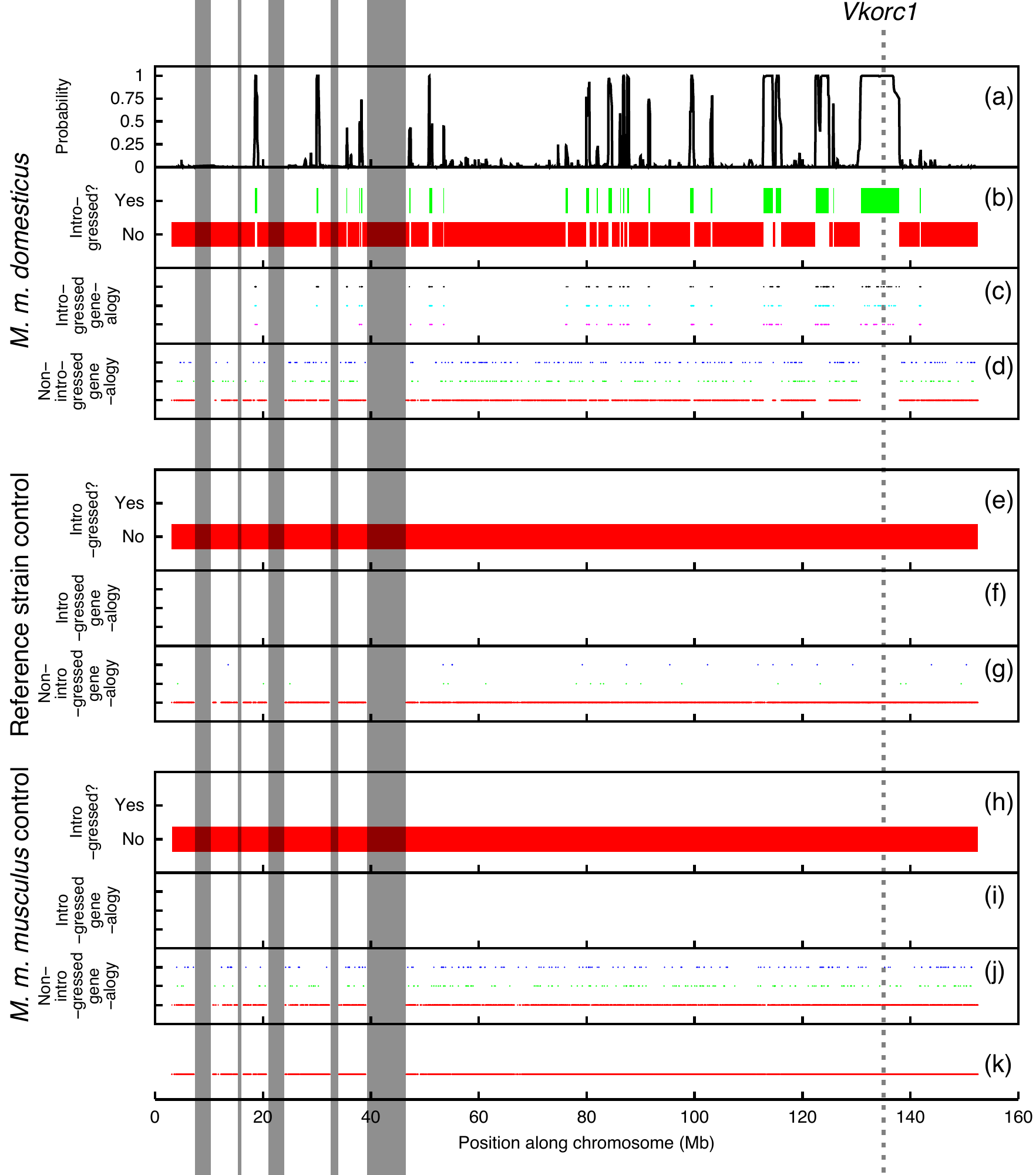}
\end{center}
\vspace{-.25in}
\caption{\small {\bf Introgression scans of chromosome 7 from three mouse genome data sets.} 
 {\it Mus musculus domesticus} samples data set  (panels b-d), 
  negative control scans of reference strains  data set (panels e-g), and 
{\em Mus musculus musculus} samples (panels h-j). 
In panels (b), (e), and (h), genomic regions are classified as having introgressed origin or not
based on parental-tree switching in a PhyloNet-HMM analysis.
Panels (c), (f), and (i) show the rooted gene genealogy inferred for each locus classified as introgressed.
Each distinct rooted gene genealogy is represented using a distinct color and row.
Panels (d), (g), and (j) show the rooted gene genealogy inferred for the remaining loci (which
were not classified as introgressed).
Panel (k) shows loci sampled by the Mouse Diversity Array \cite{Yang2009},
which we used to genotype our samples. The dashed vertical line indicates the location of the {\it Vkorc1} gene,
which was shown by \cite{Song20111296} to be a driver gene in an introgression
event between ({\it M. m. domesticus} and {\it Mus spretus}) and leading
to the spread of rodenticide resistance in the wild. The grey bars indicate regions with missing data. 
Panel (a) gives the results of posterior decoding on introgressed ($r$) states based on the results in Panel (b); see 
text for more details.
}
\label{empirical-scans}
\end{figure}

We use our new model and inference method 
to analyze two types of empirical data sets. The
first type includes individuals of known introgressed origin, and
our model recovers the introgressed genomic region reported in
\cite{Song20111296} (Fig.~\ref{empirical-scans}(b)). On the other hand, the second type consists of ``control"
individuals collected from geographically distant regions so as to
minimize the chances of introgression (though, it is not possible to
rule out that option completely). Our model detected no regions of
introgressive descent in this dataset (Fig.~\ref{empirical-scans}(e) and (h)). 

We ran PhyloNet-HMM to analyze the {\em M. m. domesticus} data set, which consisted of samples from a putative hybrid
zone between {\it M. m. domesticus} and {\it M. spretus}
(Figure \ref{empirical-scans}(b-d).
The data set contained sequences from chromosome 7, the chromosome containing
the {\it Vkorc1} gene. {\it Vkorc1} is a  gene implicated in the introgression event and
the spread of rodenticide resistance in the wild \cite{Song20111296}.

Based on the pattern of recovered parental trees,
the PhyloNet-HMM analysis detected introgression in the vicinity
of the {\it Vkorc1} gene from approximately 131 Mb to 138 Mb, 
reproducing the findings of \cite{Song20111296}.
The analysis also uncovered recombination and incomplete lineage
sorting in the region, as evidenced by incongruence among the 
rooted gene genealogies that were ascribed to loci.

The PhyloNet-HMM analysis detected introgression
in 12.0\% of sites in chromosome 7. 
Notably, the analysis located similar regions in other parts of
chromosome 7 which were not investigated by prior studies such as
\cite{Song20111296}.
Examples include the region from 122 Mb to 125 Mb and the region
from 113 Mb to 116 Mb. These introgressed regions contain about 300 genes, 
 with two groups with significant gene ontology (GO) term enrichment: one with 
 olfaction-related genes, and the other with immune-response related genes. 
It is worth mentioning that the method does detect ILS within introgressed 
regions and outside those regions as well; yet, it does not switch back and forth 
between these two cases repeatedly (which is an issue that plagues methods that 
assume independence across loci). 

As described by our model above, if we sum the transition probabilities from any $q$ state to 
all $r$ states, we obtain a value for $\gamma$. We performed this computation for each $q$ 
state, and took the average of all $\gamma$ estimates based on each of the 15 $q$ states. Our 
model estimates the value of $\gamma$ as $0.008$. This can be interpreted as the probability of 
switching due to introgression, and can shed light on introgression parameters. 

To assess confidence in these findings, we used a modified version of the posterior decoding. 
Recall that in our model, there are 15 states corresponding to the parental tree in
Fig.~\ref{phmm-parental-trees}(c): $r_1,r_2,\ldots,r_{15}$. As we are interested in assessing 
confidence in whether a column $i$ in the alignment $x$ falls within an introgressed region, 
we computed for column $i$ the quantity 
 $$p_i = \sum_{k \in \{r_1,\ldots,r_{15}\}} {\bf P}(\pi_i = k|x).$$ 
 This quantity, for all positions in chromosome 7, is shown in Fig.~\ref{empirical-scans}(a). 
 Clearly, the introgressed regions indicated by green bars in Fig.~\ref{empirical-scans}(b) have 
 very high support (close to 1), particularly the region around the {\em Vkorc1} gene. Very few 
 regions detected as introgressed by Viterbi's algorithm have low support (close to 0.25). However,
 these regions are very short. 


To further validate our approach, we repeated our scans on the reference strain control data set
and the {\em M. m. musculus} control data set, which contained two sets of
mice that were not known to hybridize (Figure \ref{empirical-scans}(e-j)).
In combination with the {\it M. spretus} samples from the previous scan,
one control data set consisted of two individuals from an inbred laboratory strain 
that were nearly genetically identical, and the other control data
set consisted of geographically and genetically distinct samples from {\it M. m. musculus}, 
which is not known to 
hybridize with {\it M. spretus} in the wild.

In both controls, PhyloNet-HMM did not detect introgression.
In the {\it M. m. musculus} control data set, the analysis recovered
signatures of recombination and ILS, based on local incongruence among inferred
rooted gene genealogies.
The scans of the laboratory strains in the reference strain control data set exhibited less local phylogenetic incongruence
compared to the scans of the wild {\it M. m. musculus} samples, as expected by the genetic
homogeneity of individuals from a single laboratory strain.


\vspace{-.2in}
\section*{Conclusion}
\vspace{-.1in}
In this paper, we introduced a new framework, PhyloNet-HMM, for comparative genomic 
analyses aimed at detecting introgression. Our framework allows for modeling point 
mutations, recombination, and introgression, and can be trained to tease apart the 
effects of incomplete lineage sorting from those of introgression. 

We implemented our model, along with standard HMM algorithms, and analyzed a data 
set of chromosome 7 from four mouse genomes where introgression was previously reported.
Our analyses detected the reported introgression with high confidence, and detected other 
regions in the chromosome as well. Using the model, we estimated that about 12\% of the sites 
in chromosome 7 in the an {\em M. m. domesticus} genome are of introgressive descent. 
Further, we ran the model on negative control data sets, and detected no introgression. 

We described above how to extend the model to general data sets with arbitrary hybridization 
and speciation events, by using a MUL-tree technique. 
 However, as larger (in terms of number of genomes) data sets become available, we expect the problem 
to become more challenging, particularly in terms of computational requirements. 
 Furthermore, while the discussion so far has assumed that the set of states is
known (equivalently, that the phylogenetic network is known), this is
not the case in practice. This is a very challenging problem that, if
not dealt with carefully, can produce poor results. In this
work, we explored a phylogenetic network corresponding to a
hypothesis provided by a practitioner. In general, the model can be ``wrapped" by a procedure 
 that iterates over all possible phylogenetic network hypotheses, and for each one the model 
 can be learned as above, and then using model selection tests, an optimal model can be selected. 
 However, this is prohibitive except for data sets with very small numbers of taxa. 
 As an alternative, the following heuristic could be adopted instead:
first, sample loci across the
genome that are distant enough to guarantee that they are unlinked;
second, use trees built on these loci to search for a
phylogenetic network topology using techniques \cite{Yu01092013}; third, 
conduct the analysis as above. Of course, the phylogenetic network
identified by the search might be inaccurate, in which
case use of an
ensemble of phylogenetic networks that are close to that one in terms
of optimality may be beneficial.

\newpage
\section*{Appendix}
\begin{table}[!ht]
\caption{{\bf Mouse samples and data sets.} }
{\footnotesize
\begin{tabular}{|c|c|c|c|c|}
\hline
{\bf Sample name} & {\bf Origin} & {\bf Gender} & {\bf Source} & {\bf Alias} \\
\hline
Spanish-mainland-domesticus & Roca del Valles, Catalunya, Spain & Female & \cite{Yang2011} & MWN1287  \\
Georgian-domesticus & Adjaria, Georgia & Male & \cite{Yang2011,Guenet200324} & DGA \\
A-spretus & Puerto Real, Cadiz Province, Spain & Male & This study & SPRET/EiJ \\
B-spretus & Sante Fe, Granada Province, Spain & Unknown & \cite{Didion2012} & SEG/Pas \\
A-reference & Classical & Male & This study & C57BL/6J \\
B-reference & Classical & Female & \cite{Didion2012} & C57BL/6J \\
A-musculus & Urumqi, Xinjiang, China & Male & \cite{Yang2011} & Yu2097m \\
B-musculus & Hebukesaier, Xinjiang, China & Female & \cite{Yang2011} & Yu2120f \\
\hline
\hline
{\bf Data set} & \multicolumn{4}{|c|}{{\bf Set of samples used}} \\
\hline
{\it M. m. domesticus} & \multicolumn{4}{|c|}{Spanish-mainland-domesticus, Georgian-domesticus, A-spretus, B-spretus} \\
Reference strain control & \multicolumn{4}{|c|}{A-reference, B-reference, A-spretus, B-spretus} \\
{\it M. m. musculus} control & \multicolumn{4}{|c|}{A-musculus, B-musculus, A-spretus, B-spretus} \\
\hline
\end{tabular}
}
\begin{flushleft}
 We obtained new mouse samples and also used existing mouse samples from
previous studies. 
The array CEL files for existing mouse samples are available online
(\verb'http://cgd.jax.org/datasets/diversityarray/CELfiles.shtml'
and by request from the authors of \cite{Didion2012}).
The introgression scans examined patterns of local phylogeny switching involving 
an
{\it M. m. domesticus} sample from the region of sympatry with
two {\it M. spretus} strains and a baseline {\it M. m. domesticus} sample
from far away.
The control scans utilized the two {\it M. spretus} strains 
along with two other mice that were known to not have introgressed with {\it M. spretus}:
either two individuals from the classical laboratory C57BL/6J strain, or
two wild {\it M. m. musculus} mice.
\end{flushleft}
\label{samples}
\end{table}

\end{document}